\documentclass[aps,prl,twocolumn,showpacs,longbibliography]{revtex4-1}
\usepackage{amstext,amsmath,amssymb,amsfonts,bbm}
\usepackage{graphicx}

\def\beq{\begin{equation}}
\def\be{\begin{equation}}
\def\ee{\end{equation}}
\def\bes{\begin{eqnarray}}
\def\ees{\end{eqnarray}}

\DeclareMathOperator{\id}{id}
\DeclareMathOperator{\SU}{SU}

\DeclareMathOperator{\Col}{Col}

\newcommand{\unit}{\mathbbm{1}}
\def\bra{\langle}
\def\ket{\rangle}

\def\pp{\partial}

\def\C{{\mathbbm C}}

\def\Z{{\mathbbm Z}}

\def\C{{\mathcal C}}

\def\H{{\mathcal H}}

\def\calM{{\mathcal M}}

\def\calZ{{\mathcal Z}}
\def\Z{{\mathcal Z}}

\usepackage[colorlinks,citecolor=blue,linkcolor=blue]{hyperref}

\begin{document}

\title{In quantum gravity, summing is refining}

\author{Carlo Rovelli}\email{rovelli@cpt.univ-mrs.fr}
\author{Matteo Smerlak}\email{smerlak@cpt.univ-mrs.fr}

\affiliation{Centre de Physique Th\'eorique, Campus de Luminy, Case 907, 13288 Marseille Cedex 09 France}

\date{\small\today}

\begin{abstract}\noindent
In perturbative QED, the approximation is improved by \emph{summing}  more Feynman graphs; in non-perturbative QCD, by \emph{refining} the lattice. Here we observe that in quantum gravity the two procedures may well be the same. We outline the combinatorial structure of spinfoam quantum gravity, define the continuum limit, and show that under general conditions refining foams is the same as summing over them. The conditions bear on the cylindrical consistency of the spinfoam amplitudes and on the presence of appropriate combinatorial factors, related to the implementation of diffeomorphisms invariance. Intuitively, the sites of the lattice are points of space: these are themselves quanta of the gravitational field, and thus a lattice discretization is also a Feynman history of quanta.
\end{abstract} 

\maketitle



A key step in constructing an interactive quantum field theory (QFT) is always  a \emph{finite} truncation of the dynamical degrees of freedom. In weakly coupled theories, such as low-energy QED or high-energy QCD, one can rely on the particle structure of the free field and consider virtual processes involving \emph{finitely many} particles, described by Feynman diagrams. In strongly coupled theories, such as confining QCD, we must resort to a non-perturbative truncation, such as a \emph{finite} lattice approximation. In either case, the relevant effect of the remaining degrees of freedom can be subsumed under a dressing of the coupling constants, if a criterion of renormalizability or criticality is met. The full theory is then formally defined by a limit where all the degrees of freedom are recovered.
The way this limit is implemented in perturbative QED and in lattice QCD, however, are quite different: in the first case, by \emph{summing} over an increasing number of Feynman graphs;  in the second case, by \emph{refining} the lattice. Here we show that in quantum gravity the two implementations of the limits may in fact turn out to be aspects of the same structure. In closure we present an intuitive physical interpretation of this result. 

It is well-known that neither the QED nor the lattice QCD scheme are available as such in quantum gravity: the perturbative expansion is non-renormalizable, and there is no background metric space to discretize. But we can still introduce a truncation of the dynamical degrees of freedom. For instance, in dynamical-triangulation approaches \cite{Ambjorn:2010rx}, discretized metric degrees of freedom are captured by fixing edge lengths of the simplices and varying their number and adjacency relations, while in Regge calculus  \cite{Regge:1961px} we fix the triangulation and vary the edge lengths. In the first case, the full dynamics is recovered by \emph{summing} over all triangulations, like one sums over Feynman graphs in perturbative QFT; in the second case,  by the limit of infinite \emph{refinement} of the triangulation, like one refines the lattice in defining the continuum limit of lattice QFT.  In loop quantum gravity (LQG), on the other hand, the discretization is induced by the diffeomorphism-invariant kinematics of gravity itself \cite{Ashtekar:1992tm}: the quantum states of space are labelled by $3$-diffeomorphism classes of spin-networks, which are discrete structures in themselves \cite{Rovelli:1995ac}. The spinfoam formalism defines \emph{truncated} transition amplitudes between two such states \cite{Rovelli:2010vv}. The amplitude is associated to a finite $2$-complex, or `foam', cobording the corresponding graphs and colored with irreducible representations of $\SU(2)$ (For a recent introduction to the covariant theory see \cite{Rovelli:2011eq}).
The question of how to recover the full set of degrees of freedom in this context has been much debated.  Should we \emph{sum} over all foams, or \emph{refine} them and take a continuum limit? Are spinfoams like Feynman diagrams and dynamical triangulations, or like metric lattices and Regge manifolds? We show below that, under suitable conditions, the two alternatives are in fact the same: the amplitude of a foam is equal to the sum of the amplitudes of all its sub-foams, with the trivial representation of $\SU(2)$ excluded from the admissible colorings.

The conditions for this to happen are of interest.  First, the vertex amplitude must obey a `cylindrical consistency' requirement.  The Euclidean vertex amplitude introduced in \cite{Engle:2007wy} and \cite{Freidel:2007py}  satisfies this condition, and so does the Lorentzian  amplitude (but not the variant proposed in \cite{Ding:2010ye}). Related conditions are studied by Magliaro and Perini \cite{Magliaro:2010ih} and by Bahr, Hellmann, Kaminski,  Kisielowski and Lewandowski \cite{Bahr:2010bs}.    Second, suitable combinatorial factors must be included in the definition of the amplitude.  These have a compelling interpretation: they correspond to quotienting by the volume of the residual discrete action of the diffeomorphism group on foams.  

\vspace{.5em}

Spinfoam models for quantum gravity can be seen as combinatorial relatives of topological quantum field theories (TQFT) \cite{Rovelli:2010vv}. In Atiyah's scheme, an $(n+1)$-dimensional TQFT is as a functorial association of a finite dimensional Hilbert  space $\H_\Sigma$ to each closed oriented $n$-manifold $\Sigma$, and a vector $\Z_\calM\in\H_{\Sigma}$ to each oriented $(n+1)$-manifold $\calM$ having $\Sigma$ as its boundary \cite{Atiyah:1988fk,Atiyah:1990uq}. Similarly, a spinfoam model associates to each oriented graph $\Gamma$ a Hilbert space $\H_\Gamma$, and to each foam $\C$ having $\Gamma$ as its boundary a vector $\calZ_\C\in\H_\Gamma$. In order to make this analogy precise, let us start with some relevant definitions. (Graphs are topological spaces as well as combinatorial objects. Likewise, foams can be defined in two ways, emphasizing their topological or their combinatorial structure. Although the former approach is often taken in the literature, after Baez \cite{Baez:1997zt,Baez:1999sr}, we find it more convenient to focus on the relevant combinatorics from scratch.) 

We call \emph{graph} a couple $\Gamma=(N_\Gamma,L_\Gamma)$ with $N_\Gamma$ a finite set of \emph{nodes} and $L_\Gamma$ a set of ordered pairs  of nodes, the \emph{links} of $\Gamma$. The nodes $n$ and $n'$ of a link $l=(n,n')$ are called the \emph{source} and \emph{target} of $l$, and denoted $s(l)$ and $t(l)$ respectively. We denote $l^{-1}\equiv(n',n)$ the reversed link, and $\overline{\Gamma}$ the graph obtained from $\Gamma$ reversing all  links.

A combinatorial two-complex, or \emph{foam}, is a triple $\C=(V_\C,E_\C, F_\C)$ with $V_\C$ a finite set of \emph{vertices}, $E_\C$ a set of ordered pairs of vertices $e=(v,v')$, now called \emph{edges}, and $F_\C$ a finite set of \emph{faces}.  Here, we call \emph{face} a finite sequence of edges $ f=(e^{\epsilon_{1}}_1,..., e^{\epsilon_{n}}_n,..., e^{\epsilon_{n_f}}_{n_f})$, where $t(e^{\epsilon_{n}}_n)=s(e^{\epsilon_{n+1}}_{n+1})$, $\epsilon_n=\pm1$ and $n_f+1:= 1$. Note that any subset $F$ of $F_{\C}$ naturally defines a subcomplex of $\C$, made of the vertices, edges and faces appearing in $F$. Finally, we denote $\overline{\C}$ the foam obtained from $\C$ by reversing all its edges and faces.

There is a notion of \emph{boundary} for foams. Indeed, let us call the edges of $E_\C$ appearing exactly once in only one face its \emph{links}, and the other ones its \emph{interior edges}. Similarly, let us call the vertices appearing exactly once in an interior edge its \emph{nodes}, and the other ones its \emph{interior vertices}. 
The sets of nodes and links of a foam $\C$ generally do not form a graph, but when they do, and moreover the orientation of each link matches the one induced by the unique face passing through it, we say that $\C$ is a \emph{proper foam}. We then define the \emph{boundary} $\pp\C$ as the subcomplex of $\C$ defined by those faces of $\C$ which contain at least one link. The underlying graph of $\pp\C$ is the \emph{boundary graph} of $\C$. In this language, foams with the same boundary graphs can still have different boundaries, because the boundary faces can be different.

We say that two proper foams $\C_1$ and $\C_2$ are \emph{composable} along an oriented graph $\Gamma$ if  $\Gamma$ is a connected component of the boundary graphs of both $\C_1$ and $\overline{\C_2}$. In this case, we can define the \emph{composition} of $\C_1$ and $\C_2$ along $\Gamma$ as the foam
$\C_1\cup_{\Gamma}\C_2$ obtained by removing $\Gamma$ and merging, for each node $n$, the unique edges $e_1^n$ and $e_2^n$ of $\C_1$ and $\C_2$ adjacent on $n$ into a single edge, and for each link $l$ the unique faces $f_1^l$ and $f_2^l$ of $\C_1$ and $\C_2$ adjacent on $l$ into a single face, see Fig. 1. Note that there is an obvious unit $\unit_\Gamma$ for the composition along $\Gamma$, the `cylinder' with exactly one face per link of $\Gamma$.

With these definitions, we can define a general spinfoam model \`a la Atiyah as the association of a Hilbert space $\H_{\Gamma}$ to each oriented graph $\Gamma$, and of a vector $\calZ_\C\in\H_{\Gamma}$ to each proper foam $\C$ with boundary graph $\Gamma$. This association should satisfy the following axioms
\\[1mm]
$\bullet$ (multiplicativity) $\H_{\Gamma_1\sqcup\Gamma_2}=\H_{\Gamma_1}\otimes\H_{\Gamma_2}$,
\\
$\bullet$  (duality) $\H_{\overline{\Gamma}}=\H_{\Gamma}^{*}\quad\textrm{and}\quad\calZ_{\overline{\C}}=\calZ_{\C}^{\dagger}$,
\\
$\bullet$  (functoriality) $\calZ_{\C_1\cup_{\Gamma}\C_2}=\langle\calZ_{\overline{\C_2}}\vert\calZ_{\C_1}\rangle_{\H_{\Gamma}}=\langle\calZ_{\overline{\C_1}}\vert\calZ_{\C_2}\rangle_{\H_{\overline{\Gamma}}}$,
\\[1mm]
to which we add the obvious axioms $\H_{\emptyset}=\mathbb{C}$ and $\calZ_{\unit_\Gamma}=\id_{\H_\Gamma}$ (using the canonical identification of elements of $\H_\Gamma\otimes\H_\Gamma^*$ with operators on $\H_\Gamma $).

\begin{figure}
\centerline{\includegraphics[scale=1]{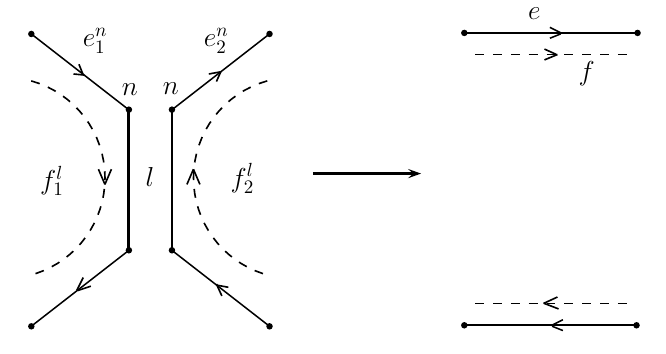}}
\caption{Gluing two foams along a link of their common boundary graph.}
\end{figure}

This set of axioms is precisely Atiyah's, except for two points. (See also Oeckl's \emph{general boundary} formulation of quantum field theory \cite{Oeckl:2005bv}.) First, because the boundary graphs are not oriented globally, like boundary manifolds, but only linkwise, the condition of functoriality cannot be framed in terms of morphisms between source and target spaces, but rather in terms of partial inner products as above. Second, we do \emph{not} require the Hilbert spaces to be finite dimensional.

The dynamics of LQG, as defined in \cite{Rovelli:2010vv,Rovelli:2011eq}, fits this framework.  In this case, the Hilbert spaces associated to the graph $\Gamma$ is the lattice $SU(2)$ Yang-Mills space $\H_\Gamma=L^2\big(SU(2)\big)^{|L_\Gamma|}/SU(2)^{|N_\Gamma|}$ (modulo the automorphisms of $\Gamma$), and the \emph{foam amplitudes} $\calZ_\C$ are defined themselves as sums $\calZ_\C=\sum_\sigma\calZ_\C(\sigma)$ of \emph{spinfoam amplitudes} $\calZ_\C(\sigma)$ over colorings $\sigma$ of the faces of $\C$ by unitary irreducible representations $j_f$ of $SU(2)$, see below.

From a physical perspective, the vectors $\psi_\Gamma\in\H_\Gamma$ are quantum states of the spatial geometry, and the inner products $\bra\Z_{\C}|\psi_{\pp\C}\ket$ the quantum amplitude, of a spacetime truncated by $\C$. To remove the dependence on $\C$, two options can be envisaged: infinitely refining $\C$, and summing over $\C$. Since the set of foams is discrete, the latter  is easy to define in principle, at least if one disregards convergence issues. But what about the former?

It can too, thanks to the following observation: the set of proper foams with a given boundary is naturally a \emph{directed set}, that is a partially ordered set where every pair of elements has an upper bound. Consider indeed the natural order relation defined by $\C_1\leq\C_2$ if there is an \emph{embedding} $\iota:\C_1\hookrightarrow\C_2$, namely a triple of injective maps $\iota_v:V_{\C_1}\rightarrow V_{\C_2}$, $\iota_e:E_{\C_1}\rightarrow E_{\C_2}$, $\iota_f:F_{\C_1}\rightarrow F_{\C_2}$ preserving the relations between vertices, edges and faces. It is easy to see that any pair of foams with the same boundary has an upper bound $\C\geq\C_1,\C_2$ with respect to this relation, given essentially by attaching the vertices, edges and faces of $\C_2$ to $\C_1$. Thus, a spinfoam model is a \emph{net} over the set of proper foams with a given boundary. The \emph{continuum limit} of $\calZ_\C$, with fixed boundary $\pp\C=c$ and boundary graph $\Gamma$, can then be defined the limit of this net in a suitable topology in $\H_{\Gamma}$, i.e. the vector $\calZ^{\infty}_{c}\in\H_{\Gamma}$ such that, for every neighborhood $\mathcal{U}$ of $\calZ^{\infty}_{c}$, there is a foam $\C_{\mathcal{U}}$ such that $\Z_{\C}\in\mathcal{U}$ whenever $\C\geq\C_{\mathcal{U}}$. 

What is the relation between the limit $\calZ^{\infty}_{c}$ and the sum $\sum_{\pp\C=c}\Z_{\C}$ (assuming both exist)? First, observe that any choice of a spinfoam amplitude $Z_{\C}(\sigma)$ defines \emph{two} spinfoam models, depending whether the trivial representations are included or not:
\be
\label{net}
\calZ_{\C}:=\!\!\!\sum_{\sigma\in\Col(\C)} \calZ_{\C}(\sigma),
 \ \ \ {\rm and} \ \ \  
\calZ^*_{\C}:=\!\!\!\sum_{\sigma\in\Col^*(\C)} \calZ_{\C}(\sigma).
\ee
Here, for a fixed foam $\C$, we denoted $\Col(\C)$ the set of all colorings $\sigma$, and $\Col^*(\C)$ the subset consisting of colorings by \emph{non-trivial} ($j_f\neq0$) representations. 

The relationship between the continuum limit $\calZ^{\infty}_{c}$ and the sum of $\calZ_\C$ over all the foams $\C$ with fixed boundary is the following: under a general condition on the amplitude $\calZ_{\C}(\sigma)$, which we detail in the next paragraph, we have
\be\label{result}\calZ_{\C}=\sum_{\substack{\C'\leq\C\\ \pp\C'=c}}\calZ^*_{\C'},
\ee
and hence, passing to the limit (assuming it exists),
\be
\calZ^{\infty}_{c} =\sum_{\pp\C=c} \calZ^*_{\C}.
\label{main}
\ee 
This is our central result. It states that refining foams in the model $\calZ$ is the same as summing over foams in the model $\Z^*$.   Note that, due to the peculiar net structure, \emph{all} foams appear in \eqref{main} in the following sense: every foam $\C$ with boundary $\pp\C=c$ appears in one finite sum \eqref{result} whose value can be chosen arbitrarily close to $\calZ^{\infty}_{c}$.

The condition on $Z_\C(\sigma)$ for the relation \eqref{main} to hold is one of \emph{cylindrical consistency}. First, observe that each coloring $\sigma$ of a foam $\C$ comes with a multiplicity, related to the symmetries of $\C$. The \emph{multiplicity} $\vert\sigma\vert_{\C}$ of a coloring $\sigma\in\Col(\C)$ is the number of colorings $\sigma'$ such that $\sigma=\sigma'\circ\phi$, with $\phi$ an automorphism of $\C$, i.e. a bijective embedding of $\C$ into itself. Then we say that the amplitude
\be
A_{\C}(\sigma) := \vert\sigma\vert_{\C}\  \calZ_{\C}(\sigma),
\label{AZ}
\ee
is \emph{cylindrically consistent} if $A_\C(\sigma)=A_{\C'}(\sigma')$ when $(\C',\sigma')$ is a \emph{trivial extension} of $(\C,\sigma)$, that is when $\C$ is a subfoam of $\C'$, $\sigma$ and $\sigma'$ coincide on the faces of $\C$ and $\sigma'$ is trivial on the other faces of $\C'$.

Under this condition, the proof of (\ref{result}) is straightforward. First, observe that the subfoams of $\C$ index a partition of $\Col(\C)$, in which each each class is made of the trivial extensions of a given subfoam $\C'\subset\C$:
 \be\Col(\C)=\bigsqcup_{\C'\subset\C}\Col^*({\C'}).\ee This implies that \be\calZ_{\C}=\sum_{\C'\subset\C}\left(\sum_{\sigma'\in\Col^*({\C'})}\vert\sigma'\vert_{\C}^{-1}A_\C(\sigma')\right).\ee Second, check that we have \be\vert\sigma'\vert_{\C}=\vert\sigma'\vert_{\C'}N_{\C',\C},\ee with $N_{\C',\C}$ the number of subfoams of $\C$ isomorphic to $\C'$. Third, use cylindrical consistency to get \be
\calZ_{\C}=\sum_{\C'\subset\C}N_{\C',\C}^{-1}\calZ^*_{\C'}
\ee
and conclude to (\ref{result}).

The LQG spinfoam amplitudes $\calZ_\C$ discussed in the literature do not satisfy the condition above.  However, a simple modification of these amplitudes does. To see this, let us start from the Lorentzian version of the LQG amplitude defined in
\cite{Engle:2007wy} and \cite{Freidel:2007py}. Using the representation given in
\cite{Rovelli:2011eq}, the amplitude of a foam with a given coloring $\sigma=(j_f)_{f\in F_\C}$ is 
\begin{eqnarray}
\calZ_{\cal C}(\sigma)&=& \int_X dG\ 
 \prod_{f}\ (2j_{\!{}_f}+1)^2
\label{sfm} \\ &&  \nonumber 
\chi^{\gamma j_f,j_f}\!\Big(H_f(G)\Big) \prod_{e\in\partial f}\chi^{j_{\!{}_f}}\!(h_{ef}(G
)).
\end{eqnarray}
where $X$ is the group $SL(2,\mathbb{C})^{n_\C}\times SU(2)^{m_\C}$ with $n_\C$ and $m_\C$ two integers depending on $\C$, $dG$ its Haar measure, $H_f:X\rightarrow SL(2,\mathbb{C})$, $h_{ef}:X\rightarrow SU(2)$, and $\chi^{\rho,j}$ (resp. $\chi^j$) characters of $SL(2,\mathbb{C})$ (resp. $SU(2)$). See \cite{Rovelli:2010vv,Rovelli:2011eq} for the rest of the definition. It is immediate to see that if $j_f=0$ then the face $f$ gives no contribution to the amplitude, since $(2j_{\!{}_f}+1)=1$ and $\chi^{0,0}=\chi^0=1$.  Hence $\calZ_{\cal C}(\sigma)$ is invariant under trivial extension. But the condition considered in the previous section was the invariance of \eqref{AZ}, not the invariance of $\calZ_{\cal C}(\sigma)$. Thus the spinfoam model \eqref{sfm} does not satisfy the condition for \eqref{main} to hold.

However, a slight modification of the model satisfies this condition. It suffices  replacing  \eqref{sfm} by 
\be
\calZ_{\cal C}(\sigma) \to  {\vert\sigma\vert_{\C}}^{-1}  \calZ_{\cal C}(\sigma)
\label{modifica}
\ee
This modification does not interfere with the interesting physical properties of the model; for instance, it does not affect any of the results on the asymptotic limit of the theory \cite{Barrett:2009cj,Conrady:2008ea,Bianchi:2009ri,Rovelli:2011uq}.  But is there any rationale for considering such a modification and expecting it to be part of the quantum theory of gravity?
There is.  The spinfoam sum is meant to be an implementation of the Misner-Hawking sum over geometries
\be
Z=\int_{\rm Metrics/Diff}\ Dg_{\mu\nu}\ e^{\frac{i}{\hbar}S[g_{\mu\nu}]}.
\label{sog}
\ee
Here the integral is not over metrics, but over equivalence classes of metrics under diffeomorphisms. In the truncation induced by the choice of a foam, the diffeomorphisms are reduced to the automorphisms $\phi$ of the foam.  Therefore the colorings $\sigma$ and $\sigma\circ\phi$ have a natural interpretation as the discrete residual equivalent of diffeomorphism-related metrics. This interpretation is reinforced by the fact that the amplitude is in fact invariant. If we want to integrate over geometries, then, the contributions of $\sigma$ and $\sigma\circ\phi$ represent an overcounting, and we must divide by the number of them, namely by $\vert\sigma\vert_{\C}$. The same conclusion can be reached by interpreting colored spinfoams as histories of nontrivial spin networks. Then colorings related by foam automorphisms clearly represent the same history. For both these reasons, it is  physically interesting to consider the amplitudes modified as in \eqref{modifica}.  Then the sum over foams is equal to the continuum limit. 

To be sure, if the multiplicities are not included, a relation between the models ${\Z}$ and ${\Z^*}$ still holds.  This is not (\ref{result}), but
\be{\Z_\C}=\sum_{\substack{\C'\leq\C\\ \pp\C'=c}}N_{\C', \C}{\Z^*_{\C'}}.\ee
That is, refining foams is still closely related to summing over foams, but less neatly. Observe also that the sum on the right-hand side is much less likely to converge in this case, as the number $N_{\C', \C}$ grows very quickly with the number of  of vertices, edges and faces of $\C $ and $\C'$.

In closing, we observe that the convergence between the perturbative-QED and the lattice-QCD pictures should not be too surprising, for \emph{it follows from the physics of general relativity}. Heuristically,  lattice sites are small regions of space; according to general relativity, these are excitations of the gravitational field, therefore they are themselves quanta of a (generally covariant) quantum field theory. This scenario is concretely realized in LQG: an $N$-quanta state of gravity has the very same structure as a Yang-Mills state on a lattice with $N$ sites  \cite{Rovelli:2011eq}. 

From this perspective, a foam is nothing but a ``history" of space quanta, in the sense in which a Feynman graph is a ``history" of field quanta.  Technically, the convergence is realized by the fact that the faces of the foam carrying vanishing spin have no effect  on the amplitude: a region where the gravitational field vanishes is like a non-existing region. This is different from lattice QCD, where a region of the lattice where the field vanishes cannot be eliminated from the lattice without altering the associated amplitude. In the latter case, if the field vanishes what remains is empty space; in the former, if the field vanishes what remains is nothing at all.  

To summarize, we have given a precise definition of the continuum limit of a spinfoam model. 
We have observed that under certain general conditions, if this limit exist, it can equally be expressed as the sum over foams, by simply restricting the amplitudes to those with nontrivial spins.  The condition for this to happen rules out a certain technical modification of the loop quantum gravity spinfoam amplitude, and suggests to modify the spinfoam amplitude by factoring away the action of the automorphisms of the foam. These can be viewed as the residual action of the diffeomorphisms group on the truncated theory.  A question we have left open is the relationship between the combinatorial coefficients introduced by this factorization and those derivied from the group field theory formulation of the theory \cite{Geloun:2010vj,Krajewski:2010yq,Baratin:2011tg}.

\vspace{.1em}

{We thank E.\ Magliaro and C.\ Perini for sending us a draft of their work,  A.\ Baratin for useful comments, J.\ Lewandowski for pointing out an error in the first version of this draft, and S.\ Speziale for helping us correcting it.}

\bibliographystyle{apsrev4-1}
\bibliography{BiblioCarlo}

\begin{thebibliography}{23}%
\makeatletter
\providecommand \@ifxundefined [1]{%
 \@ifx{#1\undefined}
}%
\providecommand \@ifnum [1]{%
 \ifnum #1\expandafter \@firstoftwo
 \else \expandafter \@secondoftwo
 \fi
}%
\providecommand \@ifx [1]{%
 \ifx #1\expandafter \@firstoftwo
 \else \expandafter \@secondoftwo
 \fi
}%
\providecommand \natexlab [1]{#1}%
\providecommand \enquote  [1]{``#1''}%
\providecommand \bibnamefont  [1]{#1}%
\providecommand \bibfnamefont [1]{#1}%
\providecommand \citenamefont [1]{#1}%
\providecommand \href@noop [0]{\@secondoftwo}%
\providecommand \href [0]{\begingroup \@sanitize@url \@href}%
\providecommand \@href[1]{\@@startlink{#1}\@@href}%
\providecommand \@@href[1]{\endgroup#1\@@endlink}%
\providecommand \@sanitize@url [0]{\catcode `\\12\catcode `\$12\catcode
  `\&12\catcode `\#12\catcode `\^12\catcode `\_12\catcode `\%12\relax}%
\providecommand \@@startlink[1]{}%
\providecommand \@@endlink[0]{}%
\providecommand \url  [0]{\begingroup\@sanitize@url \@url }%
\providecommand \@url [1]{\endgroup\@href {#1}{\urlprefix }}%
\providecommand \urlprefix  [0]{URL }%
\providecommand \Eprint [0]{\href }%
\@ifxundefined \urlstyle {%
  \providecommand \doi  [0]{\begingroup \@sanitize@url \@doi}%
  \providecommand \@doi [1]{\endgroup \@@startlink {\doibase
  #1}doi:\discretionary {}{}{}#1\@@endlink }%
}{%
  \providecommand \doi  [0]{doi:\discretionary{}{}{}\begingroup
  \urlstyle{rm}\Url }%
}%
\providecommand \doibase [0]{http://dx.doi.org/}%
\providecommand \Doi [0]{\begingroup \@sanitize@url \@Doi }%
\providecommand \@Doi  [1]{\endgroup\@@startlink{\doibase#1}\@@Doi}%
\providecommand \@@Doi [1]{#1\@@endlink}%
\providecommand \selectlanguage [0]{\@gobble}%
\providecommand \bibinfo  [0]{\@secondoftwo}%
\providecommand \bibfield  [0]{\@secondoftwo}%
\providecommand \translation [1]{[#1]}%
\providecommand \BibitemOpen [0]{}%
\providecommand \bibitemStop [0]{}%
\providecommand \bibitemNoStop [0]{.\EOS\space}%
\providecommand \EOS [0]{\spacefactor3000\relax}%
\providecommand \BibitemShut  [1]{\csname bibitem#1\endcsname}%
\bibitem [{\citenamefont {Ambjorn}\ \emph {et~al.}(2010)\citenamefont
  {Ambjorn}, \citenamefont {Jurkiewicz},\ and\ \citenamefont
  {Loll}}]{Ambjorn:2010rx}%
  \BibitemOpen
  \bibfield  {author} {\bibinfo {author} {\bibfnamefont {J.}~\bibnamefont
  {Ambjorn}}, \bibinfo {author} {\bibfnamefont {J.}~\bibnamefont {Jurkiewicz}},
  \ and\ \bibinfo {author} {\bibfnamefont {R.}~\bibnamefont {Loll}},\
  }\bibfield  {title} {\enquote {\bibinfo {title} {{Causal Dynamical
  Triangulations and the Quest for Quantum Gravity}},}\ }\href@noop {} {
  (\bibinfo {year} {2010})},\ \Eprint {http://arxiv.org/abs/1004.0352}
  {arXiv:1004.0352 [hep-th]} \BibitemShut {NoStop}%
\bibitem [{\citenamefont {Regge}(1961)}]{Regge:1961px}%
  \BibitemOpen
  \bibfield  {author} {\bibinfo {author} {\bibfnamefont {T.}~\bibnamefont
  {Regge}},\ }\bibfield  {title} {\enquote {\bibinfo {title} {{General
  relativity without coordinates}},}\ }\Doi {10.1007/BF02733251} {\bibfield
  {journal} {\bibinfo  {journal} {Nuovo Cim.},\ }\textbf {\bibinfo {volume}
  {19}},\ \bibinfo {pages} {558--571} (\bibinfo {year} {1961})}\BibitemShut
  {NoStop}%
\bibitem [{\citenamefont {Ashtekar}\ \emph {et~al.}(1992)\citenamefont
  {Ashtekar}, \citenamefont {Rovelli},\ and\ \citenamefont
  {Smolin}}]{Ashtekar:1992tm}%
  \BibitemOpen
  \bibfield  {author} {\bibinfo {author} {\bibfnamefont {Abhay}\ \bibnamefont
  {Ashtekar}}, \bibinfo {author} {\bibfnamefont {Carlo}\ \bibnamefont
  {Rovelli}}, \ and\ \bibinfo {author} {\bibfnamefont {Lee}\ \bibnamefont
  {Smolin}},\ }\bibfield  {title} {\enquote {\bibinfo {title} {{Weaving a
  classical geometry with quantum threads}},}\ }\Doi
  {10.1103/PhysRevLett.69.237} {\bibfield  {journal} {\bibinfo  {journal}
  {Phys. Rev. Lett.},\ }\textbf {\bibinfo {volume} {69}},\ \bibinfo {pages}
  {237--240} (\bibinfo {year} {1992})},\ \Eprint
  {http://arxiv.org/abs/hep-th/9203079} {arXiv:hep-th/9203079} \BibitemShut
  {NoStop}%
\bibitem [{\citenamefont {Rovelli}\ and\ \citenamefont
  {Smolin}(1995)}]{Rovelli:1995ac}%
  \BibitemOpen
  \bibfield  {author} {\bibinfo {author} {\bibfnamefont {Carlo}\ \bibnamefont
  {Rovelli}}\ and\ \bibinfo {author} {\bibfnamefont {Lee}\ \bibnamefont
  {Smolin}},\ }\bibfield  {title} {\enquote {\bibinfo {title} {{Spin networks
  and quantum gravity}},}\ }\Doi {10.1103/PhysRevD.52.5743} {\bibfield
  {journal} {\bibinfo  {journal} {Phys. Rev.},\ }\textbf {\bibinfo {volume}
  {D52}},\ \bibinfo {pages} {5743--5759} (\bibinfo {year} {1995})},\ \Eprint
  {http://arxiv.org/abs/gr-qc/9505006} {arXiv:gr-qc/9505006} \BibitemShut
  {NoStop}%
\bibitem [{\citenamefont {Rovelli}(2010)}]{Rovelli:2010vv}%
  \BibitemOpen
  \bibfield  {author} {\bibinfo {author} {\bibfnamefont {Carlo}\ \bibnamefont
  {Rovelli}},\ }\bibfield  {title} {\enquote {\bibinfo {title} {{Simple model
  for quantum general relativity from loop quantum gravity}},}\ }\href@noop {}
  { (\bibinfo {year} {2010})},\ \Eprint {http://arxiv.org/abs/1010.1939}
  {arXiv:1010.1939} \BibitemShut {NoStop}%
\bibitem [{\citenamefont {Rovelli}(2011)}]{Rovelli:2011eq}%
  \BibitemOpen
  \bibfield  {author} {\bibinfo {author} {\bibfnamefont {Carlo}\ \bibnamefont
  {Rovelli}},\ }\bibfield  {title} {\enquote {\bibinfo {title} {{Zakopane
  lectures on loop gravity}},}\ }\href@noop {} { (\bibinfo {year} {2011})},\
  \Eprint {http://arxiv.org/abs/1102.3660} {arXiv:1102.3660} \BibitemShut
  {NoStop}%
\bibitem [{\citenamefont {Engle}\ \emph {et~al.}(2008)\citenamefont {Engle},
  \citenamefont {Livine}, \citenamefont {Pereira},\ and\ \citenamefont
  {Rovelli}}]{Engle:2007wy}%
  \BibitemOpen
  \bibfield  {author} {\bibinfo {author} {\bibfnamefont {Jonathan}\
  \bibnamefont {Engle}}, \bibinfo {author} {\bibfnamefont {Etera}\ \bibnamefont
  {Livine}}, \bibinfo {author} {\bibfnamefont {Roberto}\ \bibnamefont
  {Pereira}}, \ and\ \bibinfo {author} {\bibfnamefont {Carlo}\ \bibnamefont
  {Rovelli}},\ }\bibfield  {title} {\enquote {\bibinfo {title} {{LQG vertex
  with finite Immirzi parameter}},}\ }\Doi {10.1016/j.nuclphysb.2008.02.018}
  {\bibfield  {journal} {\bibinfo  {journal} {Nucl. Phys.},\ }\textbf {\bibinfo
  {volume} {B799}},\ \bibinfo {pages} {136--149} (\bibinfo {year} {2008})},\
  \Eprint {http://arxiv.org/abs/0711.0146} {arXiv:0711.0146} \BibitemShut
  {NoStop}%
\bibitem [{\citenamefont {Freidel}\ and\ \citenamefont
  {Krasnov}(2008)}]{Freidel:2007py}%
  \BibitemOpen
  \bibfield  {author} {\bibinfo {author} {\bibfnamefont {Laurent}\ \bibnamefont
  {Freidel}}\ and\ \bibinfo {author} {\bibfnamefont {Kirill}\ \bibnamefont
  {Krasnov}},\ }\bibfield  {title} {\enquote {\bibinfo {title} {{A New Spin
  Foam Model for 4d Gravity}},}\ }\Doi {10.1088/0264-9381/25/12/125018}
  {\bibfield  {journal} {\bibinfo  {journal} {Class. Quant. Grav.},\ }\textbf
  {\bibinfo {volume} {25}},\ \bibinfo {pages} {125018} (\bibinfo {year}
  {2008})},\ \Eprint {http://arxiv.org/abs/0708.1595} {arXiv:0708.1595}
  \BibitemShut {NoStop}%
\bibitem [{\citenamefont {Ding}\ and\ \citenamefont
  {Rovelli}(2010)}]{Ding:2010ye}%
  \BibitemOpen
  \bibfield  {author} {\bibinfo {author} {\bibfnamefont {You}\ \bibnamefont
  {Ding}}\ and\ \bibinfo {author} {\bibfnamefont {Carlo}\ \bibnamefont
  {Rovelli}},\ }\bibfield  {title} {\enquote {\bibinfo {title} {{Physical
  boundary Hilbert space and volume operator in the Lorentzian new spin-foam
  theory}},}\ }\Doi {10.1088/0264-9381/27/20/205003} {\bibfield  {journal}
  {\bibinfo  {journal} {Class. Quant. Grav.},\ }\textbf {\bibinfo {volume}
  {27}},\ \bibinfo {pages} {205003} (\bibinfo {year} {2010})},\ \Eprint
  {http://arxiv.org/abs/1006.1294} {arXiv:1006.1294} \BibitemShut {NoStop}%
\bibitem [{\citenamefont {Magliaro}\ and\ \citenamefont
  {Perini}(2010)}]{Magliaro:2010ih}%
  \BibitemOpen
  \bibfield  {author} {\bibinfo {author} {\bibfnamefont {Elena}\ \bibnamefont
  {Magliaro}}\ and\ \bibinfo {author} {\bibfnamefont {Claudio}\ \bibnamefont
  {Perini}},\ }\bibfield  {title} {\enquote {\bibinfo {title} {{Local spin
  foams}},}\ }\href@noop {} { (\bibinfo {year} {2010})},\ \Eprint
  {http://arxiv.org/abs/1010.5227} {arXiv:1010.5227} \BibitemShut {NoStop}%
\bibitem [{\citenamefont {Bahr}\ \emph {et~al.}(2011)\citenamefont {Bahr},
  \citenamefont {Hellmann}, \citenamefont {Kaminski}, \citenamefont
  {Kisielowski},\ and\ \citenamefont {Lewandowski}}]{Bahr:2010bs}%
  \BibitemOpen
  \bibfield  {author} {\bibinfo {author} {\bibfnamefont {Benjamin}\
  \bibnamefont {Bahr}}, \bibinfo {author} {\bibfnamefont {Frank}\ \bibnamefont
  {Hellmann}}, \bibinfo {author} {\bibfnamefont {Wojciech}\ \bibnamefont
  {Kaminski}}, \bibinfo {author} {\bibfnamefont {Marcin}\ \bibnamefont
  {Kisielowski}}, \ and\ \bibinfo {author} {\bibfnamefont {Jerzy}\ \bibnamefont
  {Lewandowski}},\ }\bibfield  {title} {\enquote {\bibinfo {title} {{Operator
  Spin Foam Models}},}\ }\Doi {10.1088/0264-9381/28/10/105003} {\bibfield
  {journal} {\bibinfo  {journal} {Class. Quant. Grav.},\ }\textbf {\bibinfo
  {volume} {28}},\ \bibinfo {pages} {105003} (\bibinfo {year} {2011})},\
  \Eprint {http://arxiv.org/abs/1010.4787} {arXiv:1010.4787 [gr-qc]}
  \BibitemShut {NoStop}%
\bibitem [{\citenamefont {Atiyah}(1988)}]{Atiyah:1988fk}%
  \BibitemOpen
  \bibfield  {author} {\bibinfo {author} {\bibfnamefont {Michael}\ \bibnamefont
  {Atiyah}},\ }\href@noop {} {\emph {\bibinfo {title} {Topological quantum
  field theory}}},\ Vol.~\bibinfo {volume} {68}\ (\bibinfo  {publisher}
  {Publication mathematiques de l'I.H.E.S.},\ \bibinfo {year}
  {1988})\BibitemShut {NoStop}%
\bibitem [{\citenamefont {Atiyah}(1990)}]{Atiyah:1990uq}%
  \BibitemOpen
  \bibfield  {author} {\bibinfo {author} {\bibfnamefont {Michael}\ \bibnamefont
  {Atiyah}},\ }\href@noop {} {\emph {\bibinfo {title} {The geometry and
  phyisics of knots}}}\ (\bibinfo  {publisher} {Cambridge University Press},\
  \bibinfo {year} {1990})\BibitemShut {NoStop}%
\bibitem [{\citenamefont {Baez}(1998)}]{Baez:1997zt}%
  \BibitemOpen
  \bibfield  {author} {\bibinfo {author} {\bibfnamefont {John~C.}\ \bibnamefont
  {Baez}},\ }\bibfield  {title} {\enquote {\bibinfo {title} {{Spin foam
  models}},}\ }\Doi {10.1088/0264-9381/15/7/004} {\bibfield  {journal}
  {\bibinfo  {journal} {Class. Quant. Grav.},\ }\textbf {\bibinfo {volume}
  {15}},\ \bibinfo {pages} {1827--1858} (\bibinfo {year} {1998})},\ \Eprint
  {http://arxiv.org/abs/gr-qc/9709052} {arXiv:gr-qc/9709052} \BibitemShut
  {NoStop}%
\bibitem [{\citenamefont {Baez}(2000)}]{Baez:1999sr}%
  \BibitemOpen
  \bibfield  {author} {\bibinfo {author} {\bibfnamefont {John~C.}\ \bibnamefont
  {Baez}},\ }\bibfield  {title} {\enquote {\bibinfo {title} {{An introduction
  to spin foam models of BF theory and quantum gravity}},}\ }\href@noop {}
  {\bibfield  {journal} {\bibinfo  {journal} {Lect. Notes Phys.},\ }\textbf
  {\bibinfo {volume} {543}},\ \bibinfo {pages} {25--94} (\bibinfo {year}
  {2000})},\ \Eprint {http://arxiv.org/abs/gr-qc/9905087} {arXiv:gr-qc/9905087}
  \BibitemShut {NoStop}%
\bibitem [{\citenamefont {Oeckl}(2008)}]{Oeckl:2005bv}%
  \BibitemOpen
  \bibfield  {author} {\bibinfo {author} {\bibfnamefont {Robert}\ \bibnamefont
  {Oeckl}},\ }\bibfield  {title} {\enquote {\bibinfo {title} {{General boundary
  quantum field theory: Foundations and probability interpretation}},}\
  }\href@noop {} {\bibfield  {journal} {\bibinfo  {journal} {Adv. Theor. Math.
  Phys.},\ }\textbf {\bibinfo {volume} {12}},\ \bibinfo {pages} {319--352}
  (\bibinfo {year} {2008})},\ \Eprint {http://arxiv.org/abs/hep-th/0509122}
  {arXiv:hep-th/0509122} \BibitemShut {NoStop}%
\bibitem [{\citenamefont {Barrett}\ \emph {et~al.}(2009)\citenamefont
  {Barrett}, \citenamefont {Dowdall}, \citenamefont {Fairbairn}, \citenamefont
  {Gomes},\ and\ \citenamefont {Hellmann}}]{Barrett:2009cj}%
  \BibitemOpen
  \bibfield  {author} {\bibinfo {author} {\bibfnamefont {John~W.}\ \bibnamefont
  {Barrett}}, \bibinfo {author} {\bibfnamefont {Richard~J.}\ \bibnamefont
  {Dowdall}}, \bibinfo {author} {\bibfnamefont {Winston~J.}\ \bibnamefont
  {Fairbairn}}, \bibinfo {author} {\bibfnamefont {Henrique}\ \bibnamefont
  {Gomes}}, \ and\ \bibinfo {author} {\bibfnamefont {Frank}\ \bibnamefont
  {Hellmann}},\ }\bibfield  {title} {\enquote {\bibinfo {title} {{A Summary of
  the asymptotic analysis for the EPRL amplitude}},}\ }\href@noop {} {
  (\bibinfo {year} {2009})},\ \Eprint {http://arxiv.org/abs/0909.1882}
  {arXiv:0909.1882} \BibitemShut {NoStop}%
\bibitem [{\citenamefont {Conrady}\ and\ \citenamefont
  {Freidel}(2008)}]{Conrady:2008ea}%
  \BibitemOpen
  \bibfield  {author} {\bibinfo {author} {\bibfnamefont {Florian}\ \bibnamefont
  {Conrady}}\ and\ \bibinfo {author} {\bibfnamefont {Laurent}\ \bibnamefont
  {Freidel}},\ }\bibfield  {title} {\enquote {\bibinfo {title} {{Path integral
  representation of spin foam models of 4d gravity}},}\ }\Doi
  {10.1088/0264-9381/25/24/245010} {\bibfield  {journal} {\bibinfo  {journal}
  {Class. Quant. Grav.},\ }\textbf {\bibinfo {volume} {25}},\ \bibinfo {pages}
  {245010} (\bibinfo {year} {2008})},\ \Eprint {http://arxiv.org/abs/0806.4640}
  {arXiv:0806.4640} \BibitemShut {NoStop}%
\bibitem [{\citenamefont {Bianchi}\ \emph {et~al.}(2009)\citenamefont
  {Bianchi}, \citenamefont {Magliaro},\ and\ \citenamefont
  {Perini}}]{Bianchi:2009ri}%
  \BibitemOpen
  \bibfield  {author} {\bibinfo {author} {\bibfnamefont {Eugenio}\ \bibnamefont
  {Bianchi}}, \bibinfo {author} {\bibfnamefont {Elena}\ \bibnamefont
  {Magliaro}}, \ and\ \bibinfo {author} {\bibfnamefont {Claudio}\ \bibnamefont
  {Perini}},\ }\bibfield  {title} {\enquote {\bibinfo {title} {{LQG propagator
  from the new spin foams}},}\ }\Doi {10.1016/j.nuclphysb.2009.07.016}
  {\bibfield  {journal} {\bibinfo  {journal} {Nucl. Phys.},\ }\textbf {\bibinfo
  {volume} {B822}},\ \bibinfo {pages} {245--269} (\bibinfo {year} {2009})},\
  \Eprint {http://arxiv.org/abs/0905.4082} {arXiv:0905.4082} \BibitemShut
  {NoStop}%
\bibitem [{\citenamefont {Rovelli}\ and\ \citenamefont
  {Zhang}(2011)}]{Rovelli:2011uq}%
  \BibitemOpen
  \bibfield  {author} {\bibinfo {author} {\bibfnamefont {Carlo}\ \bibnamefont
  {Rovelli}}\ and\ \bibinfo {author} {\bibfnamefont {Mingyi}\ \bibnamefont
  {Zhang}},\ }\href@noop {} {\enquote {\bibinfo {title} {Euclidean three-point
  function in loop and perturbative gravity},}\ } (\bibinfo {year} {2011}),\
  \bibinfo {note} {to appear}\BibitemShut {NoStop}%
\bibitem [{\citenamefont {Geloun}\ \emph {et~al.}(2010)\citenamefont {Geloun},
  \citenamefont {Gurau},\ and\ \citenamefont {Rivasseau}}]{Geloun:2010vj}%
  \BibitemOpen
  \bibfield  {author} {\bibinfo {author} {\bibfnamefont {Joseph~Ben}\
  \bibnamefont {Geloun}}, \bibinfo {author} {\bibfnamefont {Razvan}\
  \bibnamefont {Gurau}}, \ and\ \bibinfo {author} {\bibfnamefont {Vincent}\
  \bibnamefont {Rivasseau}},\ }\bibfield  {title} {\enquote {\bibinfo {title}
  {{EPRL/FK Group Field Theory}},}\ }\href@noop {} { (\bibinfo {year}
  {2010})},\ \Eprint {http://arxiv.org/abs/1008.0354} {arXiv:1008.0354}
  \BibitemShut {NoStop}%
\bibitem [{\citenamefont {Krajewski}\ \emph {et~al.}(2010)\citenamefont
  {Krajewski}, \citenamefont {Magnen}, \citenamefont {Rivasseau}, \citenamefont
  {Tanasa},\ and\ \citenamefont {Vitale}}]{Krajewski:2010yq}%
  \BibitemOpen
  \bibfield  {author} {\bibinfo {author} {\bibfnamefont {Thomas}\ \bibnamefont
  {Krajewski}}, \bibinfo {author} {\bibfnamefont {Jacques}\ \bibnamefont
  {Magnen}}, \bibinfo {author} {\bibfnamefont {Vincent}\ \bibnamefont
  {Rivasseau}}, \bibinfo {author} {\bibfnamefont {Adrian}\ \bibnamefont
  {Tanasa}}, \ and\ \bibinfo {author} {\bibfnamefont {Patrizia}\ \bibnamefont
  {Vitale}},\ }\bibfield  {title} {\enquote {\bibinfo {title} {{Quantum
  corrections in the Group Field Theory formulation of the EPRL/FK models}},}\
  }\href@noop {} { (\bibinfo {year} {2010})},\ \Eprint
  {http://arxiv.org/abs/1007.3150} {arXiv:1007.3150} \BibitemShut {NoStop}%
\bibitem [{\citenamefont {Baratin}\ \emph {et~al.}(2011)\citenamefont
  {Baratin}, \citenamefont {Girelli},\ and\ \citenamefont
  {Oriti}}]{Baratin:2011tg}%
  \BibitemOpen
  \bibfield  {author} {\bibinfo {author} {\bibfnamefont {Aristide}\
  \bibnamefont {Baratin}}, \bibinfo {author} {\bibfnamefont {Florian}\
  \bibnamefont {Girelli}}, \ and\ \bibinfo {author} {\bibfnamefont {Daniele}\
  \bibnamefont {Oriti}},\ }\bibfield  {title} {\enquote {\bibinfo {title}
  {{Diffeomorphisms in group field theories}},}\ }\href@noop {} { (\bibinfo
  {year} {2011})},\ \Eprint {http://arxiv.org/abs/1101.0590} {arXiv:1101.0590}
  \BibitemShut {NoStop}%
\end{thebibliography}%
\end{document}